\documentclass[journal=jacsat,manuscript=article]{achemso}

\usepackage{chemformula} % Formula subscripts using \ch{}
\usepackage[T1]{fontenc} % Use modern font encodings
\usepackage{physics,graphicx}
\usepackage{mathrsfs,bm}
\author{Mingran Zhang}
\affiliation{Department of Physics, City University of Hong Kong, Kowloon, 999077 Hong Kong SAR.}

\author{Jiahao Joel Fan}
\affiliation{Department of Physics, City University of Hong Kong, Kowloon, 999077 Hong Kong SAR.}

\author{Frank Schlawin}
\affiliation{Max Planck Institute for the Structure and Dynamics of Matter,
Center for Free Electron Laser Science, Luruper Chaussee 149, 22761 Hamburg, Germany}
\alsoaffiliation{University of Hamburg, Luruper Chaussee 149, 22761 Hamburg, Germany}
\email{frank.schlawin@uni-hamburg.de}

\author{Zhedong Zhang}
\email{zzhan26@cityu.edu.hk}
\affiliation{Department of Physics, City University of Hong Kong, Kowloon, 999077 Hong Kong SAR.}
\alsoaffiliation{Shenzhen Research Institute, City University of Hong Kong, Shenzhen, Guangdong 518057, China}

\title[]
  {Theory for Entangled-Photons Stimulated Raman Scattering versus Nonlinear Absorption for Polyatomic Molecules}

\begin{document}

\begin{tocentry}

\centering
    \includegraphics[width=0.75\textwidth]{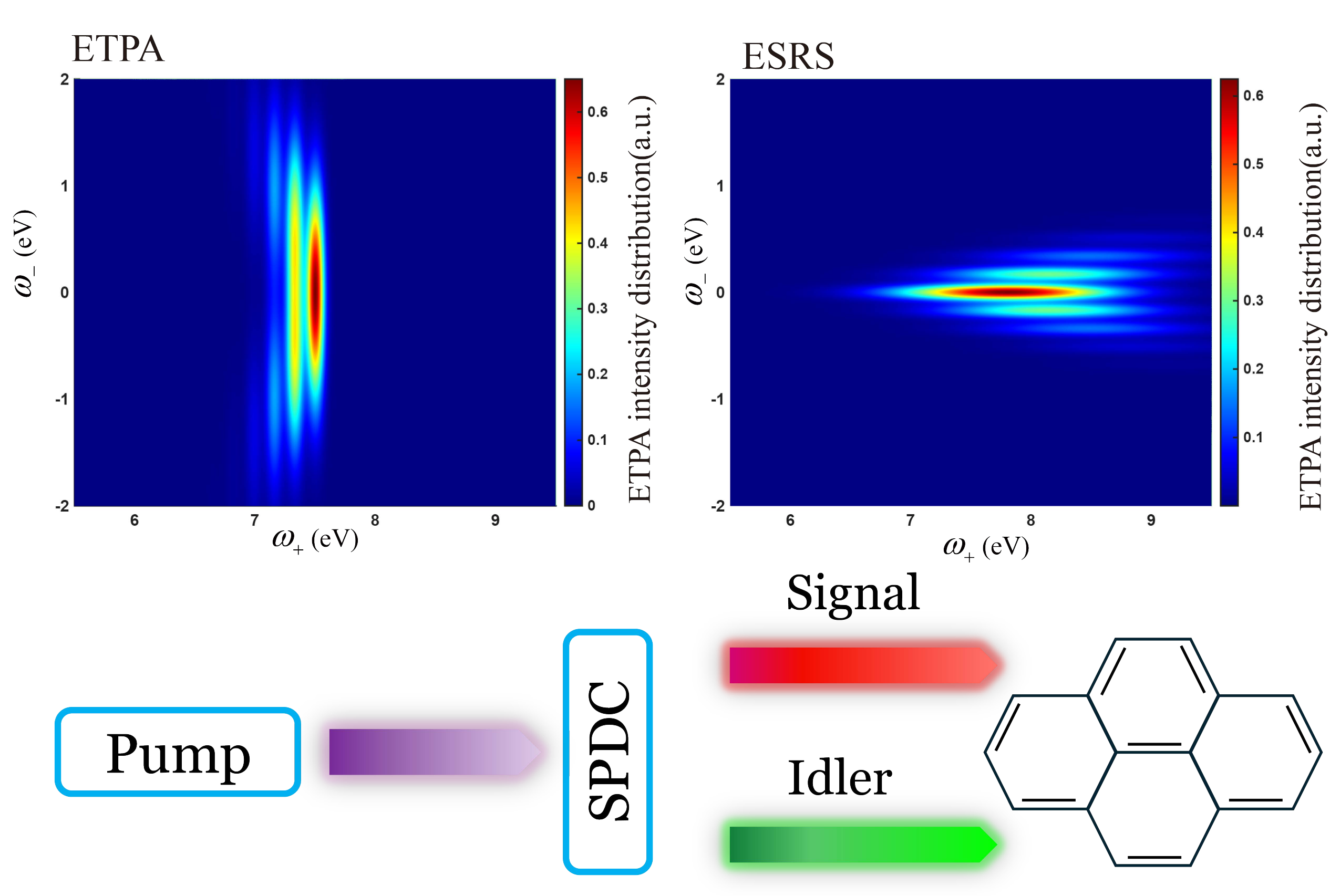}

\end{tocentry}

\begin{abstract}
Quantum entanglement offers an incredible resource for enhancing the sensing and spectroscopic probes. Here we develop a microscopic theory for the stimulated Raman scattering (SRS) using entangled photons. We demonstrate that the time-energy correlation of the photon pairs can optimize the signal for polyatomic molecules. Our results show that the spectral-line intensity of the entangled-photon SRS (ESRS) is of the same order of magnitude as the one for the entangled two-photon absorption (ETPA); the parameter window is thus identified to do so. Moreover, the vibrational coherence is found to play an important role for enhancing the ESRS against the ETPA intensity. Our work paves a firm road for extending the schemes of molecular spectroscopy with quantum light, based on the observation of the ETPA in experiments.
\end{abstract}

Nonlinear optics is essential for investigating ultrafast dynamics in complex quantum systems ranging from atoms to biological tissues, and for chemically selective and subdiffraction imaging. As two representative third-order nonlinear optical effects, two-photon absorption (TPA) and stimulated Raman scattering (SRS) have become invaluable in the development of ultrafast spectroscopy\cite{95--PrinciplesNonlinearOptical,08-PRA-Nonlinearopticalspectroscopy,09-PRA-Multidimensionalpumpprobe,22-AoCR-EntangledPhotonSpectroscopy,18-AoCR-EntangledTwoPhoton,16-RoMP-Nonlinearopticalsignals} and microstructure dynamics\cite{96-TJoCP-Collectivecoordinatesnuclear,14-TJoPCL-StimulatedRamanSpectroscopy,22-LS&A-Entangledphotonsenabled,22-AP-MonitoringWavepacketDynamics}. They are also acting as powerful tools in imaging\cite{24-P-StimulatedRamanScattering}, optical storage\cite{14-LS&A-Opticalstoragearrays}, and sensing\cite{15-PRL-AtomLightHybrid}. Applications based on classical light sources are limited by the shot noise. Thus, theoretical works focus on quantum light sources to tackle this challenge. Entangled photon pairs, which can be generated by parametric down conversion\cite{08--NonlinearOptics,11-RoSI-InvitedReviewArticle}, have been widely utilized in quantum information\cite{20-L&PR-QuantumKeyDistribution,18-PRL-18QubitEntanglement}, quantum computation\cite{15-S-Universallinearoptics,19-nQI-controlledNOTgate,20-N-Programmablephotoniccircuits}, linear quantum imaging\cite{24-NP-Advancesquantumimaging,25-BOE-Applicationquantumimaging} and quantum sensing\cite{21-PRA-Multimodemetrologyvia,21-JoMP-Multiparameterquantummetrology,21-PRA-Improvedphasesensitivity,19-O-Experimentalmultiphaseestimation}.\par

Entangled two-photon absorption (ETPA) offers significant advantages over classical TPA. Owing to its linear dependence on photon flux, in contrast to the quadratic dependence in classical TPA \cite{90-PRA-Linearintensitydependence,95-PRL-NonclassicalExcitationAtoms,06-TJoPCB-EntangledPhotonAbsorption,97-PRL-EntanglementInducedTwo}, ETPA enables the probing of biological samples with reduced risk of photodamage \cite{18-AoCR-EntangledTwoPhoton,23-PotNAoS-Colorsentangledtwo}. Furthermore, the unique quantum correlations inherent in entangled photons facilitate novel spectroscopic capabilities, such as access to classical forbidden transitions \cite{04-PRL-InducingDisallowedTwo,16-RoMP-Nonlinearopticalsignals} and pathway-selective control of light–matter interactions \cite{18-AoCR-EntangledTwoPhoton,20-APL-Interferometrictwophoton,21-SA-Distinguishability“whichpathway”,24-TJoCP-Pathwayselectivitytime}. These properties have inspired various experimental implementations in molecular detection and dynamics studies \cite{23-PotNAoS-Colorsentangledtwo,24-TJoCP-Pathwayselectivitytime,22-TJoPCA-ExperimentalStudyValidity,20-TJoPCC-MeasurementsEntangledTwo,18-JotACS-TwoPhotonExcitation}, highlighting its promising potential in quantum spectroscopy. There was a debate recently on the performance of ETPA versus TPA with lasers, between experimental and theoretical groups\cite{20-TJoPCC-MeasurementsEntangledTwo,23-PotNAoS-Colorsentangledtwo,21-OE-Quantifyingenhancementtwo,22-PRA-Aspectstwophoton,24-PRA-Limitationsfluorescencedetected}. Many researchers have tried ETPA experiments without finding the evidence of linear dependence on photon flux\cite{24-PRA-Limitationsfluorescencedetected,21-OE-Quantifyingenhancementtwo}, this could be caused by the low signal intensity compared with the environmental noise\cite{13-JoPBAMaOP-Photonstatisticsintense,25-PRL-TwoDimensionalElectronic}. The ETPA could still be improved under proper spatial and temporal conditions\cite{25-OQ-Enhancingupconversionspace–time}.  \par

As another prominent third-order nonlinear process, entangled stimulated Raman scattering (ESRS) has also garnered considerable interest. Studies indicate that ESRS can achieve signal enhancements beyond classical SRS \cite{14-TJoPCL-StimulatedRamanSpectroscopy,22-LS&A-Entangledphotonsenabled,21-PRR-EnhancingstimulatedRaman,25--Ultimateresolutionlimits,25--TheoryQuantumEnhanced},  which provide future applications in molecular vibration sensing, live-cell imaging, chemical mapping, and drug delivery monitoring \cite{22-LS&A-Entangledphotonsenabled,24-P-StimulatedRamanScattering}. However, while ESRS has been demonstrated experimentally using squeezed light sources \cite{20-O-Quantumenhancedcontinuous,21-N-Quantumenhancednonlinear}, it has not yet been realized with entangled photon pairs. Such sources are experimentally more controllable and easily applied to theoretical modeling \cite{21-SA-Distinguishability“whichpathway”,22-TJoCP-Entangledtwophoton,24-TJoPCA-EntangledTwoPhoton}, yet predictive guidelines for achievable ESRS signal levels under realistic conditions remain elusive.\par

\begin{figure}[t]
            \centering
            \includegraphics[width=0.9\textwidth]{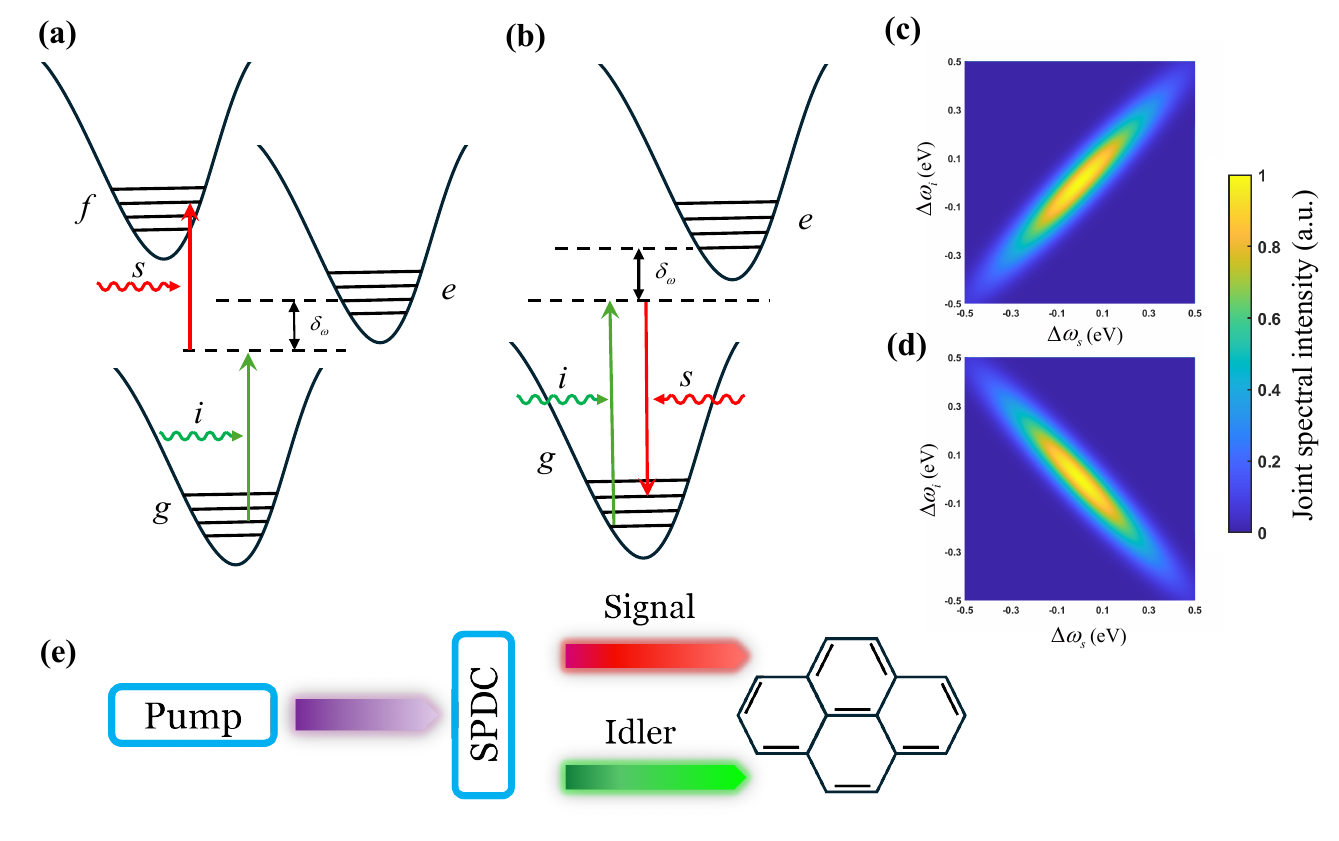}
            \caption{Level scheme of molecular system. $g,~e,~f$ represents different electron levels, the group of solid black lines are vibrational modes, $\delta_\omega$ between the dashed lines represents the detuning between real energy state and virtual touched one. (a) ETPA process, the molecule absorbs idler photon first to the $e$ state, and then absorbs signal photon to the final $f$ state. (b) ESRS process, the molecule absorbs idler photon first to the $e$ state, and then stimulated emits a photon with the same frequency of signal photon back to the $g$ state. (c) Joint spectral intensity $|F_{PDC}|^2$ of entangled photon pair. Correlated photons $\Omega_m=0.05\text{fs}^{-1},~\Omega_p=0.3\text{fs}^{-1}$ for ESRS. (d) Frequency anti-correlated photon pairs with $\Omega_p=0.05\text{fs}^{-1},~\Omega_m=0.3\text{fs}^{-1}$ for ETPA. (e) Experimental setup for the proposed ETPA and ESRS measurement. A pump photon is down-converted into two photons with frequencies $\omega_s$ and $\omega_i$. Both photons are then directed to the molecule sample. }\label{fig:re24:molecular_scheme}
        \end{figure}

As a theoretical illustration, we numerically compare the signal strengths of ETPA and ESRS to evaluate the feasibility of observing SRS with entangled photon pairs. To have a simple picture, we first consider a three-energy level system with transition frequencies $\omega_{eg}$ and $\omega_{fe}$ between the ground state $\ket{g}$, the first excited state $\ket{e}$ and the final state $\ket{f}$ for TPA process. Likewise, the three levels are represented by $\ket{g_1}$, $\ket{g_2}$ representing two lower states, and one virtual upper state $\ket{e}$ with transition frequencies $\omega_{eg_1}$ and $\omega_{eg_2}$. A level scheme is shown in Fig.~\ref{fig:re24:molecular_scheme}(a) and \ref{fig:re24:molecular_scheme}(b).

The parametric down conversion (PDC) process can generate twin-entangled photons as the form\cite{14-TJoPCL-StimulatedRamanSpectroscopy}
\begin{equation}
  |\psi_{\text{ent}}\rangle=\iint_{-\infty}^{+\infty}\dd \omega_s\dd \omega_i F_{PDC}(\omega_s,\omega_i)a^\dagger(\omega_s)a^\dagger(\omega_i)\ket{0}.
\end{equation}
Here $s$ and $i$ represent the signal and ideal photon separately, $\omega_j$ denotes the frequency corresponding to the photon $j$. The preparation of entangled photon pairs is shown in Fig.~\ref{fig:re24:molecular_scheme}(e). The function $F_{PDC}$ is the two-photon wave function which describes the behavior of generated entangled photon pairs
\begin{equation}\label{equ-spectum-frequency}
  F_{PDC}(\omega_s,\omega_i)=\frac{1}{\sqrt{\pi\Omega_m\Omega_p}}f(\frac{\Omega_s+\Omega_i}{2\Omega_p})f(\frac{\Omega_s-\Omega_i}{2\Omega_m}).
\end{equation}
Here
\begin{equation}
  f(x)=e^{-x^2}
\end{equation}
is the typical Gaussian distribution.
The above $\Omega_k=\omega_k-\omega_k^{(0)}$ is the detuning with respect to the central frequency $\omega_k^{(0)}$. $\Omega_p$ is the coefficient corresponding to the frequency sum bandwidth while $\Omega_m$ determines to the frequency difference bandwidth. \par

To provide a meaningful comparison between the behavior of two processes, we employ both frequency-correlated and frequency-anti-correlated entangled photon pairs. As shown in Equ.~\ref{equ-spectum-frequency}, a smaller value of $\Omega_p$ results in a narrower bandwidth for $\omega_+=\omega_{s}^{(0)}+\omega_{i}^{(0)}$, indicating that the sum of the input photons' central frequencies becomes the dominant factor. Such photon pairs with spectral broadening are referred to as anti-correlated, which corresponds to the case of the ETPA signal in Fig.~\ref{fig:re24:molecular_scheme}(c).

For the ESRS process, the difference coefficient $\Omega_p$ is set to a higher value so that $\omega_-=\omega_{s}^{(0)}-\omega_{i}^{(0)}$ becomes the dominant variable. This type of photon pair is termed positively correlated, as illustrated in Fig.~\ref{fig:re24:molecular_scheme}(d). Standard PDC process can generate anti-correlated photon pairs due to the energy conservation, which is convenient for ETPA experiments. Frequency correlated photon pairs can be realized by extra time lens\cite{16-PRL-SpectrallyEngineeringPhotonic}, changing
the pump pulse or collected modes characteristics\cite{16-PRA-Spectralcorrelationcontrol}, or counter-propagating phase-matching\cite{21-SR-Observationfrequencyuncorrelated}. In experiments, the correlated photons can be generated through the combination of nonlinear interference and phase control devices. After the photon pairs have been generated from the nonlinear crystals in the nonlinear interference with PDC, the spectral function would then be modulated by the phase control devices\cite{24-LSA-Entangledphotonsenabled}. 
        \par

The full three-level system can be described by
\begin{equation}
  \hat{H}_{\text{atom}}=\hbar\omega_g\op{g}{g}+\hbar\omega_e\op{e}{e}+\hbar\omega_f\op{f}{f}.
\end{equation}
with  the interaction Hamiltonian $H_{I}$ under the rotating wave approximation
\begin{equation}
  \hat{H}_{I}=-\hat{V}^\dagger(t)\cdot \hat{E}(t)+h.c.
\end{equation}
Here $\hat{V}$ is the transition dipole moment operator of the form
\begin{equation}
    \hat{V}^{\dagger}(t)=\mu_{eg}e^{i\omega_{eg}t}\op{e}{g}+\mu_{fe}e^{i\omega_{fe}t}\op{f}{e}.
\end{equation}
The electrical field operator has the form
\begin{equation}
  \hat{E}(t)=\Gamma \int \dd \omega \hat{a}(\omega)e^{-i\omega t}.
\end{equation}
Here, $\omega$ is the frequency related to the field.\par

The probability of the TPA process is proportional to the modulus square of the probability amplitude $T_{fi}$.
\begin{align}
 T_{fi} & = \bra{0}\otimes\bra{f}\ket{\psi_{\text{final}}}\notag\\
  & \approx \Gamma_s\Gamma_i\sqrt{\frac{1}{\pi\Omega_m\Omega_p}}\qty(-\frac{i}{\hbar})^2\int_{-\infty}^{+\infty}\int_{-\infty}^{\tau_1}\dd \tau_2\dd \tau_1 \bra{0}\otimes\bra{f}\hat{H}_{I}(\tau_2)\hat{H}_{I}(\tau_1)\ket{g}\otimes\ket{\psi_{\text{ent}}}\notag
  %&=\Gamma_{\text{all}}\sqrt{\frac{1}{\pi\Omega_m\Omega_p}}\int_{-\infty}^{+\infty}\int_{-\infty}^{\tau_1}\dd \tau_2\dd \tau_1\iint\dd\omega_s\dd\omega_ie^{i\omega_{fe}\tau_2}e^{i\omega_{eg}\tau_1}e^{-i\tau_2\omega_s}e^{-i\tau_1\omega_i} f\left(\frac{\Omega_s+\Omega_i}{2\Omega_p} \right) f\left(\frac{\Omega_s-\Omega_i}{2\Omega_m} \right),
\end{align}
with  $\Gamma_{\text{all}}=\Gamma_{s}\Gamma_{i}\mu_{ge}\mu_{ef}/\hbar^2$. The integral above can be evaluated analytically, i.e.,
\begin{equation}
    T_{fi}=2\pi\Gamma_{\text{all}}e^{iz/c(\omega_{fe}+\omega_{eg})} f(\frac{\omega_{fe}+\omega_{eg}-\omega_+}{2\Omega_p})[-i\pi e^{-\frac{(\omega_{\text{all}}^{\text{TPA}})^2}{\Omega_m^2}}+\mathcal{P}\int_{-\infty}^{+\infty}\frac{e^{-\frac{(\omega_i-\omega_{\text{all}}^{\text{TPA}})^2}{\Omega_m^2}}}{\omega_i-\omega_{eg}}\dd\omega_i]
\label{Tfi}
\end{equation}
where  $\omega_{\text{all}}^{\text{TPA}}=1/2(\omega_i^{(0)}+\omega_{fe}-\omega_{eg}-\omega_s^{(0)})$ is a frequency coefficient. 
The Cauchy principal value can be evaluated, i.e.,
\begin{equation}
  \mathcal{P}\int_{-\infty}^{+\infty}\frac{e^{-\frac{(\omega_i-\omega_{\text{all}})^2}{\Omega_m^2}}}{\omega_i-\omega_{eg}}\dd\omega_i\approx e^{-\frac{\omega_{\text{all}^2}}{\Omega_m^2}}(\frac{2\omega_{\text{all}}}{\Omega_m})^2
\label{Pint}
\end{equation}
with the details given in Supplemental Information.
Hence using Eqs.(\ref{Tfi},\ref{Pint}), the transition rates can be calculated, i.e., for the TPA process
\begin{equation}
  P_{\text{TPA}}=\abs{T_{fi}}^2\propto f\left(\frac{\omega_{fe}+\omega_{eg}-\omega_+}{\sqrt{2}\Omega_p} \right) f\left(\frac{\sqrt{2}\omega_{\text{all}}^{\text{TPA}}}{\Omega_m} \right)
\end{equation}
and for the SRS process
\begin{equation}
  P_{\text{SRS}}\propto f\left(\frac{\omega_{eg_2}-\omega_{eg_1}-\omega_-}{\sqrt{2}\Omega_m} \right) f\left(\frac{\sqrt{2}\omega_{\text{all}}^{\text{SRS}}}{\Omega_p} \right), \quad \omega_{\text{all}}^{\text{SRS}} = \frac{1}{2 \big(\omega_i^{(0)}+\omega_s^{(0)}-\omega_{eg_2}-\omega_{eg_1} \big)}.
\end{equation}

Thus the probability ratio of two processes is
\begin{equation}
\begin{aligned}
  &\frac{P_{\text{TPA}}}{P_{\text{SRS}}}=\mathop{\frac{\Gamma_{\text{TPA}}^2}{\Gamma_{\text{SRS}}^2}}\mathop{\frac{\Omega_m^{\text{SRS}}\Omega_p^{\text{SRS}}}{\Omega_m^{\text{TPA}}\Omega_p^{\text{TPA}}}}\\
  \times &\mathop{\exp\qty(\frac{-(\omega_{fe}+\omega_{eg}-\omega_+)^2}{2(\Omega_{p}^{\text{TPA}})^2}+\frac{(\omega_{eg^{(2)}}-\omega_{eg^{(1)}}+\omega_-)^2}{2(\Omega_{m}^{\text{SRS}})^2})}\\
\times &\mathop{\exp\qty(-\frac{(\Delta \omega_{eg}-\Delta\omega_{fe})^2}{2(\Omega_m^{\text{TPA}})^2}+\frac{(\Delta \omega_{eg_1}+\Delta\omega_{eg_2})^2}{2(\Omega_p^{\text{SRS}})^2})}\label{q-atomic-comparison}
\end{aligned}
\end{equation}
Here $\Gamma_{\text{TPA}}$ is $\Gamma_{\text{all}}$ for ETPA process and so is $\Gamma_{\text{SRS}}$.
\par
Eq.(\ref{q-atomic-comparison}) provides a comprehensive framework for comparing the probabilities of the two nonlinear processes. The first term, $\Gamma$, corresponds to the dipole moment interaction. The ratio of $\Omega$, determines a spectral broadening coefficient. The final two exponential terms are associated with the total detuning and the differential detuning, respectively. For a given material, Eq.(\ref{q-atomic-comparison}) allows direct assessment of whether the ESRS signal is comparable to that of ETPA when using a specific input photon source.

For the $\Gamma$ term,
\begin{align}
  \Gamma_{\text{TPA}}&=\Gamma_s\Gamma_i\mu_{ge}\mu_{ef}\qty(-\frac{i}{\hbar})^2\\
  \Gamma_{\text{SRS}}&=\Gamma_s\Gamma_i \mu_{g_2e}\mu_{eg_1}\qty(-\frac{i}{\hbar})^2
\end{align}
and
\begin{equation}
  \Gamma_j=\frac{i}{(2\pi)^{3/2}}\qty(\frac{\hbar\omega_{0j}}{2\epsilon_0 c})^{1/2}
\end{equation}
for $j=s$ or $i$.
Thus
\begin{equation}
  \frac{\Gamma_{\text{TPA}}^2}{\Gamma_{\text{SRS}}^2}\approx \qty|\frac{\mu_{ge}\mu_{ef}}{\mu_{eg_2}\mu_{eg_1}}|^2.
\end{equation}
which is mainly decided by the material properties.\par
The final comparison reveals that the overall strength difference between ETPA and ESRS is primarily influenced by three factors: frequency detuning, transition dipole moments, and the bandwidth of the entangled photon-pair. This suggests that by exercising precise control over the frequency characteristics of the light source and selecting materials with comparable dipole moments for the relevant transitions, it is possible to achieve an ESRS signal intensity on par with that of ETPA. Many materials exhibiting similar transition dipole strengths for $\omega_{eg}$ and $\omega_{fe}$ are thus suitable candidates for both types of experiments\cite{04-CJoC-Theoreticalstudiesone,10-TCA-Theoreticalstudyone,22-DaP-Influenceelectronwithdrawing,20-CPL-Photophysicalproperties}. However, it should be noted that the present model is solely applicable within a three-level framework.\par

The three-level atomic model provides a simplified theoretical framework that does not account for molecular effects, e.g., excited-state relaxation and dynamics. In reality, however, interactions between ions within molecules give rise to numerous vibrational modes, which split single electronic energy levels into multiple closely spaced sub-levels. To more accurately simulate potential experimental outcomes, we incorporate an additional harmonic oscillator bath to model the vibrational dynamics, as illustrated in Fig. \ref{fig:re24:molecular_scheme}.  \par

We consider a three-band electronic system that is diagonally coupled to the harmonic oscillator bath, as described by the Hamiltonian:\cite{04-CR-ManyBodyApproaches}
\begin{equation}
\begin{aligned}
\hat{H}=&\sum_b\Omega_b\op{b}+\hat{H}_{I}+\hat{H}_{SB}.\\
  \hat{H}_{SB}=&\sum_a\hat{Q}_{a}\dyad{a},\\
  \hat{Q}_a=&\sum_j c_{aj}\hat{q}_j.
  \end{aligned}
\end{equation}
Here $b$ is the index for the electronic state, the three terms represent the molecular Hamiltonian, the interaction with the optical field, and the interaction with the bath. $\hat{Q}_a$ is the collective coordinate which introduces fluctuations into the system.\par

The harmonic oscillator model is a foundational and simplified theoretical construct for simulating molecular dynamics. As a widely applied model for over three decades to successfully capture common spectroscopic features, such as the positions of absorption and fluorescence peaks as well as the general pathways of vibronic relaxation. It has the limitation like the reactivity varying between molecules. Nevertheless, for the specific purpose of analyzing ultrafast spectroscopic signals, this model provides an invaluable conceptual and analytical framework. It provides a general picture of key signal attributions, including peak positions and spectral bandwidth evolution. Thus, it enables a unified theoretical understanding of how molecular systems interact with quantum light.
\par

Similarly to the three-level model, the ESRS signal corresponds to the population change of the ground state $g$. The electron is excited by an idler photon and subsequently undergoes emission stimulated by the signal photon\cite{95--PrinciplesNonlinearOptical}.
\begin{equation}
  P_{\text{SRS}}=\int [\dd t] F(t_4,t_3,t_2,t_1)\expval{E^{\dagger}(t_4)E(t_3)E^{\dagger}(t_2)E(t_1)}
\end{equation}
Here $[\dd t]=\dd t_1\dd t_2\dd t_3\dd t_4$, $F$ is the third-order response function
\begin{equation}
\begin{aligned}
  &F(t_4,t_3,t_2,t_1)\equiv \expval{\hat{V}(t_4)\hat{V}^{\dagger}(t_3)\hat{V}(t_2)\hat{V}^{\dagger}(t_1)}\\
  &=\sum_{abcd}\rho_{aa}\ev{\hat{V}_{ad}(t_4)\hat{V}^{\dagger}_{dc}(t_3)\hat{V}_{cb}(t_2)\hat{V}^{\dagger}_{ba}(t_1)}.
  \end{aligned}
\end{equation}
under the electron eigenstate basis. $\hat{V}_{ij}$ is the transition dipole operator between the state $i$ and $j$.\par
This fourth-order correlation function can be simplified by the second order cumulant expansion\cite{22-AP-MonitoringWavepacketDynamics,09-CR-CoherentMultidimensionalOptical}
\begin{equation}
\begin{aligned}
  &\ev{\hat{V}_{ad}(t_4)\hat{V}^{\dagger}_{dc}(t_3)\hat{V}_{cb}(t_2)\hat{V}^{\dagger}_{ba}(t_1)}=\mu_{ad}\mu_{dc}\mu_{cb}\mu_{ba}\\
  &\times\exp[\phi_{cbag}+i\qty(\omega_{ad}t_4+\omega_{dc}t_3+\omega_{cb}t_2+\omega_{ba}t_1)]
\end{aligned}
\end{equation}
where
\begin{equation}
\begin{aligned}
  &\phi_{cbag}=-g_{cc}(\tau_{43})-g_{bb}(\tau_{32})-g_{aa}(\tau_{21})-g_{cb}(\tau_{42})\\
  &+g_{cb}(\tau_{43})+g_{cb}(\tau_{32})-g_{ca}(\tau_{41})+g_{ca}(\tau_{41})+g_{ca}(\tau_{42})\\
  &+g_{ca}(\tau_{31})-g_{ca}(\tau_{32})-g_{ba}(\tau_{31})+g_{ba}(\tau_{32})+g_{ba}(\tau_{21})
\end{aligned}
\end{equation}
Here $\tau_{ij}=t_i-t_j$ and
\begin{equation}
  g_{ab}(\tau)\equiv \int_0^{\tau}\dd\tau_1\int_0^{\tau_1}\dd\tau_2\ev{\hat{Q}_a(\tau_1)\hat{Q}_b(\tau_2)}
\end{equation}
is the lineshape function.
In our case, $\phi_{cbag}=\phi_{egeg}$ for ESRS signal. $g_{ab}$ has the exact form
\begin{equation}
  g_{ab}(t)=\lambda^2\qty[\coth\qty(\frac{\beta\hbar\omega_j}{2})(1-\cos(\omega_j t))+i\qty(\sin(\omega_j t)-\omega_j t)]
\end{equation}
Here $\omega_j$ is the frequency corresponding to the vibrational mode. $\lambda=1/2d_j^2$ is the Franck-Condon factor of representing the coupling strength of the nuclear to the electronic transition. $d_j$ is the dimensionless displacement of the equilibrium configuration of the vibrational mode $j$.
With the high temperature limit, this expression can be further simplified as
\begin{equation}
  g_{ab}(t)=g_{ab,l}(t)+g_{ab,h}(t)
\end{equation}
with the low (high) frequency vibrational mode
\begin{subequations}
\begin{align}
  g_{ab,l}(t)= & -\frac{k_BT\omega_j}{\hbar}D_{ab}t^2\\
  g_{ab,h}(t)= & F_{ab}\qty(1-e^{-i\omega_j t})
  \end{align}
\end{subequations}
where $D_{ab}$ ($F_{ab}$) is the fluctuation between system-bath coupling respect to low (high) frequency mode.\par
The above derivations lead to
\begin{equation}
  \phi_{egeg}=-\frac{k_BT\omega_j}{\hbar}Du^2-F_j(1-e^{i\omega_j u})
\end{equation}
Here $u=t_1-t_3$. The correlation function of the electrical field further gives
\begin{equation}
\begin{aligned}
  &\expval{E^{\dagger}(t_4)E(t_3)E^{\dagger}(t_2)E(t_1)}=\frac{\Gamma_{s}^2\Gamma_{i}^2}{\pi\Omega_m\Omega_p}\int\dd\omega_s\int \dd\omega_i\\
  &e^{-i\qty(t_1+t_4-\frac{2z}{c})\omega_i}e^{i\qty(t_2+t_3-\frac{2z}{c})\omega_s}f(\frac{\Omega_s+\Omega_i}{2\Omega_p})f(\frac{\Omega_s-\Omega_i}{2\Omega_m})
\end{aligned}
\end{equation}
which has the final representation as

\begin{equation}\label{equ-molecular-SRS}
\begin{aligned}
  &P_{\text{SRS}}\propto\sum_n\qty(e^{-F_j}\frac{F_j^n}{n!})f(\frac{\omega_--n\omega_j}{\sqrt{2}\Omega_m})\int \dd\omega\int_0^{\infty}\dd t_1 \exp\qty[i(2\omega_i)t_1]\\
  &\int_{-\infty}^{t_1}\dd u \exp\qty(-\tilde{D}_ju^2-i(\omega_jn+\omega_{eg})u)f(\frac{\omega-\omega_{\text{all}}}{\Omega_p})
  \end{aligned}
\end{equation}
Here $u=t_1-t_3$, $\tilde{D}_j=k_BT\omega_jD_j/\hbar$, and $\omega_{\text{all}}=1/2(\omega_i^{(0)}+\omega_s^{(0)}+n\omega_j)$ .\par
The ETPA process can be evaluated similarly
\begin{equation}
P_{\text{TPA}}=\int [\dd t] \expval{\hat{V}(t_4)\hat{V}(t_3)\hat{V}^{\dagger}(t_2)\hat{V}^{\dagger}(t_1)}\expval{E^{\dagger}(t_4)E^{\dagger}(t_3)E(t_2)E(t_1)}
\end{equation}
We have $\phi_{cbag}=\phi_{efeg}$ for ETPA signal. By further neglecting the low frequency vibrations due to the fast interaction interval, we have
\begin{equation}
  \phi_{efeg}=F_j\qty(e^{i\omega_j(t_3-t_1)}-1).
\end{equation}
The derived integral has the form
\begin{equation}\label{equ-molecular-TPA}
\begin{aligned}
  &P_{\text{TPA}}\propto\sum_n\qty(e^{-F_j}\frac{F_j^n}{n!})f(\frac{\omega_{fg}-\omega_+-n\omega_j}{2\Omega_p})\int\dd\omega\\
  &\int_0^{\infty}\dd t_1\int_0^{\infty}\dd t_3 f(\frac{\omega-\omega_{\text{all}}}{\Omega_m})e^{i(t_1-t_3)\omega}e^{-i(t_1-t_3)\omega_{eg}}
  \end{aligned}
\end{equation}

The functions Equ. \ref{equ-molecular-SRS} and Equ. \ref{equ-molecular-TPA} can be further simplified to achieve
\begin{equation}\label{equ-molecular-SRS-final}
\begin{aligned}
  &P_{\text{SRS}}\propto\sum_nS_nf(\frac{\omega_--n\omega_j}{\Omega_m})\frac{\sqrt{\pi}}{(n\omega_j+\omega_{eg})\tilde{D}_j}f(\frac{n\omega_j-\delta_{\text{SRS}}}{\sqrt{2}\Omega_p})
  \end{aligned}
\end{equation}
\begin{equation}\label{equ-molecular-TPA-final}
  \begin{aligned}
    P_{\text{TPA}}\propto\sum_nS_nf(\frac{\omega_+-\omega_{fg}+n\omega_j}{\sqrt{2}\Omega_p})f(\frac{n\omega_j-\delta_{\text{TPA}}}{\sqrt{2}\Omega_m})
  \end{aligned}
\end{equation}
where $S_n=\qty(e^{-F}\frac{F^n}{n!})$ is the Franck-Condon factor\cite{26-TotFS-Elementaryprocessesphotochemical,50-PotRSoLSAMaPS-TheoryLightAbsorption}.\par

\begin{figure}[ht]
\centering
\includegraphics[width=0.8\textwidth]{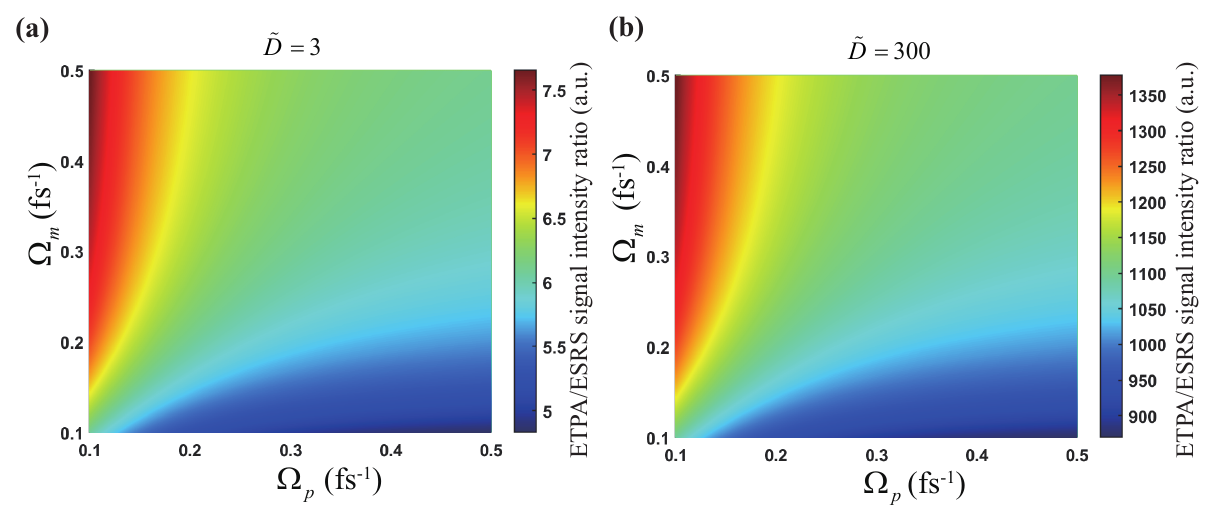}
\caption{TPA/SRS ratio with the same input entangled photon pairs. $\tilde{D}$ stands for a global decay rate of high-frequency vibrations, as induced by low-frequency vibrational modes. (a) $\tilde{D}=3$ as a comparable result for the probability of ETPA and ESRS. (b) $\tilde{D}=300$ case when ETPA signal is way more stronger than ESRS.}
\label{fig:re24:equal_omega_D}
\end{figure}\par

%\newpage
Using our theoretical model, we simulate the signal intensities for each process and compare their behavior under various conditions. The pyrene molecule is selected as the sample system, with transition dipole moments of $\mu_{eg} = 4.35~\text{D}$ and $\mu_{fe} = 6.99~\text{D}$ for the transitions between the levels $g$, $e$ and $f$ \cite{10-TCA-Theoreticalstudyone}. The energy gap $\omega_{eg} = 3.9~\text{eV}$ corresponds to a wavelength of $317~\text{nm}$, and $\omega_{fe} = 3.6~\text{eV}$ corresponds to $341~\text{nm}$. The vibrational mode parameters are adopted from pyrene calculations \cite{21-C-HowPredictExcited}, with a primary mode characterized by a Huang–Rhys factor $F = 1$ and frequency $\omega_j = 0.17~\text{eV}$. This work considers all simulations and analyses under room-temperature conditions.

Figure~\ref{fig:re24:equal_omega_D} shows the ratio between ETPA and ESRS signals in pyrene under identical incident entangled photon-pair conditions. When the photon pairs are frequency anticorrelated, corresponding to larger values of $\Omega_m$ (upper left triangular region), ETPA dominates over ESRS. In contrast, ESRS intensity increases in the lower right triangular region, where photon pairs exhibit frequency correlation as $\Omega_p$ increases. The dimensionless parameter $\tilde{D}$ represents the decay rate in Eq.~\ref{equ-molecular-SRS}. As shown in Fig.~\ref{fig:re24:equal_omega_D}(b), the signal ratio is two orders of magnitude larger than in Fig.~\ref{fig:re24:equal_omega_D}(a). Due to the fast interaction interval, the ETPA process does not contain the low-frequency decay term, thus the magnitude difference fact indicats that ESRS strength diminishes with increasing $\tilde{D}$. This suggests that materials with weaker low-frequency vibrational coupling may enhance the ESRS signal. In our simulations, the decay rate for pyrene is $\tilde{D} = 0.536$, under which the signals of both processes become comparable under specific pumping conditions. Commonly used fluorescence molecules usually have the decay rate around 1, thus would not introduce orders of magnitude difference in ESRS signal strength.

\begin{figure}[t]
\centering
\includegraphics[width=0.9\textwidth]{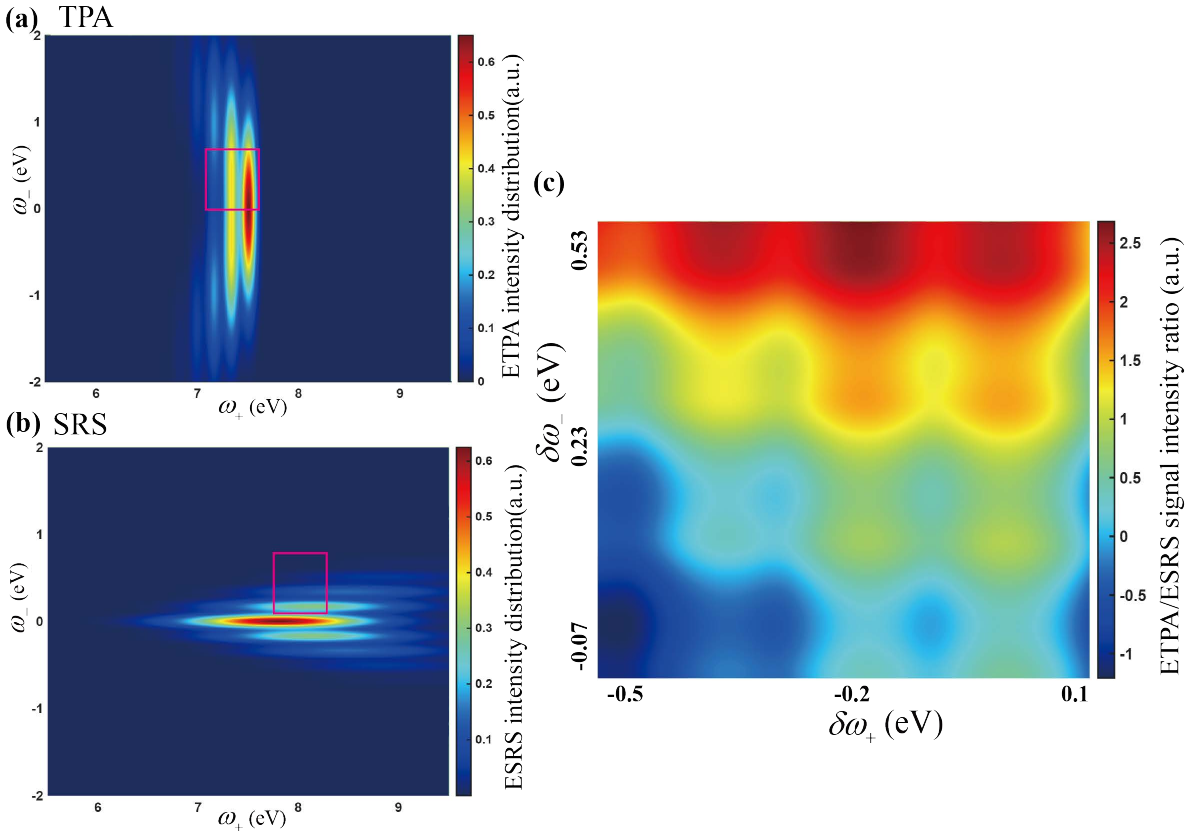}
\caption{The signal strength of ETPA, ESRS and their comparison under central frequency sum and difference. (a) Signal distribution for ETPA. The central peak is at $\omega_+=7.5\text{eV},~\omega_-=0\text{eV}$. (b) Signal distribution for ESRS. The first SRS signal peak is at $\omega_+=8.14\text{eV},~\omega_-=\pm 0.17\text{eV}$. (c) The ratio of ETPA over ESRS in the magenta rectangular region. $\delta\omega_+$ and $\delta\omega_-$ are the relative frequency compared with the central peak of the signals. ($\delta\omega_+^{\text{TPA}}, \delta\omega_-^{\text{TPA}}$)=($\omega_+^{\text{TPA}}-7.5$eV,$\omega_+^{\text{TPA}}$), ($\delta\omega_+^{\text{SRS}},\delta\omega_-^{\text{SRS}}$)=($\omega_+^{\text{SRS}}-8.14$eV, $\omega_+^{\text{SRS}}-0.17$eV). The image shows the logarithmic value.}
\label{fig:re24:omega+_omega-}
\end{figure}\par

To quantitatively evaluate the signal strength difference, we select anti-correlated photon pairs for the ETPA process with $\Omega_p = 0.05 \text{fs}^{-1},~\Omega_m = 0.3\text{fs}^{-1}$, and correlated entangled photons for ESRS with $\Omega_p = 0.3\text{fs}^{-1},~\Omega_m = 0.05\text{fs}^{-1}$, as illustrated in Fig.~\ref{fig:re24:molecular_scheme}. Fig.~\ref{fig:re24:omega+_omega-} shows the variation in signal intensity as a function of the sum and difference of the central frequencies, $\omega+$ and $\omega_-$. As shown in Fig.~\ref{fig:re24:omega+_omega-}(a) and (b), ETPA is more sensitive to changes in $\omega+$, while ESRS is strongly influenced by $\omega_-$. This behavior can be understood from the global detuning terms $f\left(\frac{\omega_- - n\omega_j}{\Omega_m}\right)$ in Equation~\ref{equ-molecular-SRS-final} and $f\left(\frac{\omega_+ - \omega_{fg} + n\omega_j}{\sqrt{2}\Omega_p}\right)$ in Equation~\ref{equ-molecular-TPA-final}.

It should be noted that the central spectral positions of the two signals do not coincide. For ESRS, the peak at $\omega_+ = 8.14~\text{eV}$ is close to the sum $\omega_{eg_1} + \omega_{eg_2} = 7.97~\text{eV}$, the discrepancy attributed to detuning effects and contributions from multiple Raman resonances. The dominant peak in Fig.~\ref{fig:re24:omega+_omega-}(b) corresponds to the Rayleigh frequency, which does not alter the frequencies of the input and output photons. For a further comparison between the highest power strengths of ESRS and ETPA, we pick the signals within the magenta rectangular regions indicated in both panels. A judgement of their experimental performances thus being exhibited. The resulting ratio of ETPA to ESRS signals is presented in Fig.~\ref{fig:re24:omega+_omega-}(c). Since the peaks of two processes do not match, we use $\delta\omega_+$ and $\delta\omega_-$ to illustrate the relative frequency difference compared with the central peak of the signals. For example, $\delta\omega_+^{\text{TPA}}=\omega_+^{\text{TPA}}-7.5$eV and $\delta\omega_-^{\text{SRS}}=\omega_-^{\text{SRS}}-0.17$eV. The represented peaks correspond to different vibrational modes, which can not be shown in three level system.

A comparison of the peak intensities in Fig.~\ref{fig:re24:omega+_omega-}(a) and (b) yields a ratio of $5.60$, indicating that the maximum signal strengths are relatively close. The broader spectral features are associated with various vibrational modes, as represented by the $n$ -th terms in Equations~\ref{equ-molecular-SRS-final} and \ref{equ-molecular-TPA-final}.

These results demonstrate that, under suitable conditions, ETPA and ESRS signals can achieve comparable intensities with appropriately chosen input photon frequencies. However, the central frequencies for the two processes differ, implying that the photon pair must be tailored specifically for either ETPA or ESRS. Alternatively, a fixed input photon pair may be used to achieve comparable signal power within the green region without significant intensity loss. Thus, Fig.~\ref{fig:re24:omega+_omega-} further supports the conclusion that, in environments with similar detuning conditions, the signal strengths of ETPA and ESRS can be made comparable.

\begin{figure}[t!]
\centering
\includegraphics[width=0.8\textwidth]{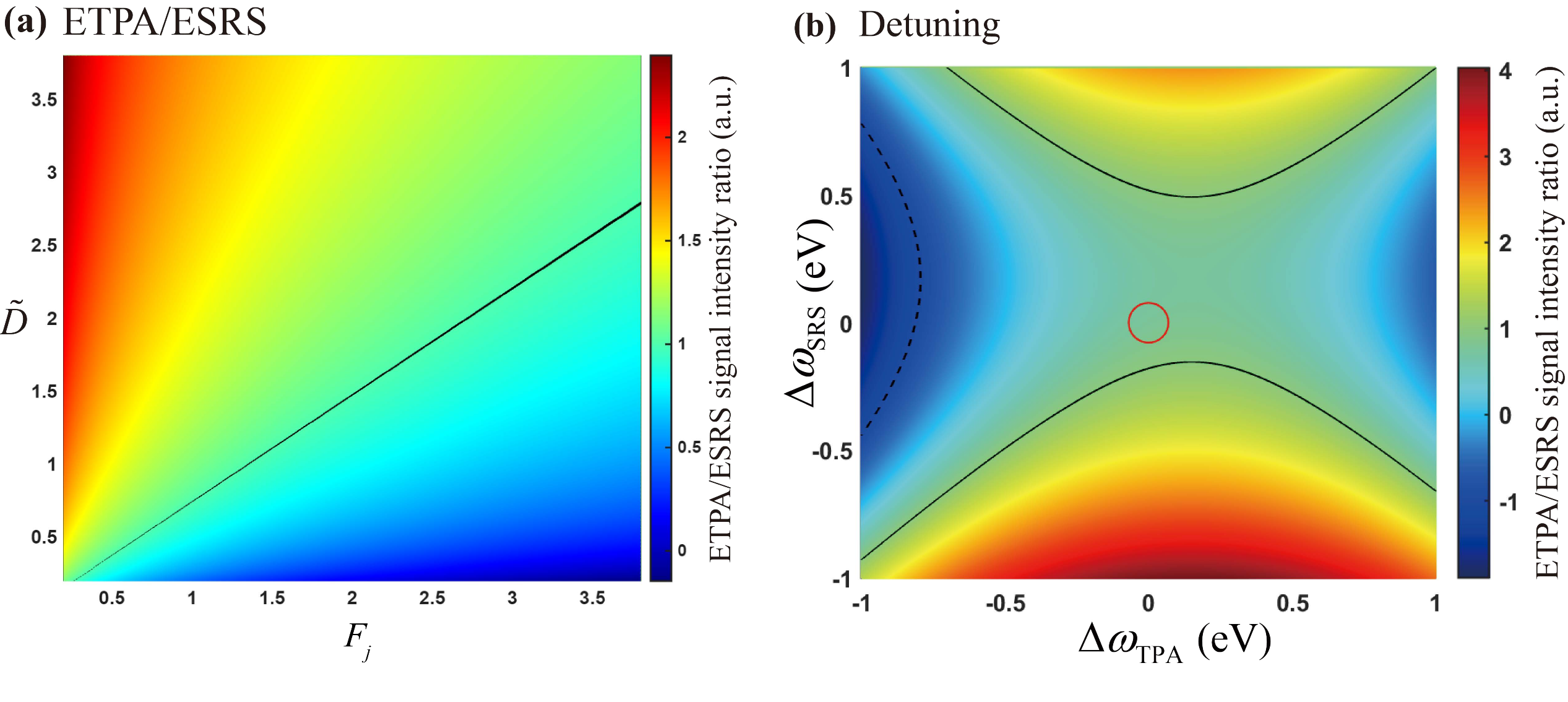}
\caption{(a) ETPA/ESRS intensity via high frequency coupling strength $F_j$ and low frequency decay rate $\tilde{D}$. The Black solid line represents the conditions where ETPA signal is an order of magnitude higher than ESRS.(b) ETPA and ESRS signal strength via the change of middle state detunings. The x-label is the variety of detuning for ETPA, and the y-label is the change of detuning for ESRS. The black solid line is the curve when ETPA is tenfold of ESRS, and the black dashed line describes when ESRS is tenfold of ETPA. The central red circle represents the region where both detunings are small. Both images show the logarithmic value. }
\label{fig:re24:detuning}
\end{figure}\par
Depart from the three level model, molecular system contains the effect from high frequency and low frequency vibrations. The relationship between Huang-Rhys factor $F_j$, low frequency decay rate $\tilde{D}$ and the signal intensity is shown in Fig.~\ref{fig:re24:detuning}(a). Although a molecule with large $F_j$ is not convenient for ETPA experiment, it might still satisfy the requirements for ESRS with small $\tilde{D}$. This is the unique advantage for molecular model to find novel experimental friendly particles.

To gain deeper understanding of the effect of intermediate-state detuning, we provide a detailed analysis in Fig.~\ref{fig:re24:detuning}(b). The central frequencies of the input photon pairs are fixed to eliminate final-state detuning. Under these conditions, variations in the electronic energy levels correspond directly to detunings between different transitions.

At the central point within the red circle, where both detunings are zero, the ratio of ETPA to ESRS signal intensities remains approximately constant at $6.10$, indicating that the two processes exhibit signal strengths of the same order of magnitude.

The black dashed and solid lines in Fig.~\ref{fig:re24:detuning} highlight the region where the signal intensities of both processes are comparable. This region primarily spans $\Delta\omega_{\text{TPA}} = \pm 0.5$ and $\Delta\omega_{\text{SRS}} = \pm 0.5$, suggesting a broad tolerance for intermediate-state detuning in both nonlinear processes. We notice that the signal intensities are not perfectly symmetric with respect to the detuning, which can be attributed to contributions from other vibrational states.

\begin{figure}[t!]
\centering
\includegraphics[width=0.6\textwidth]{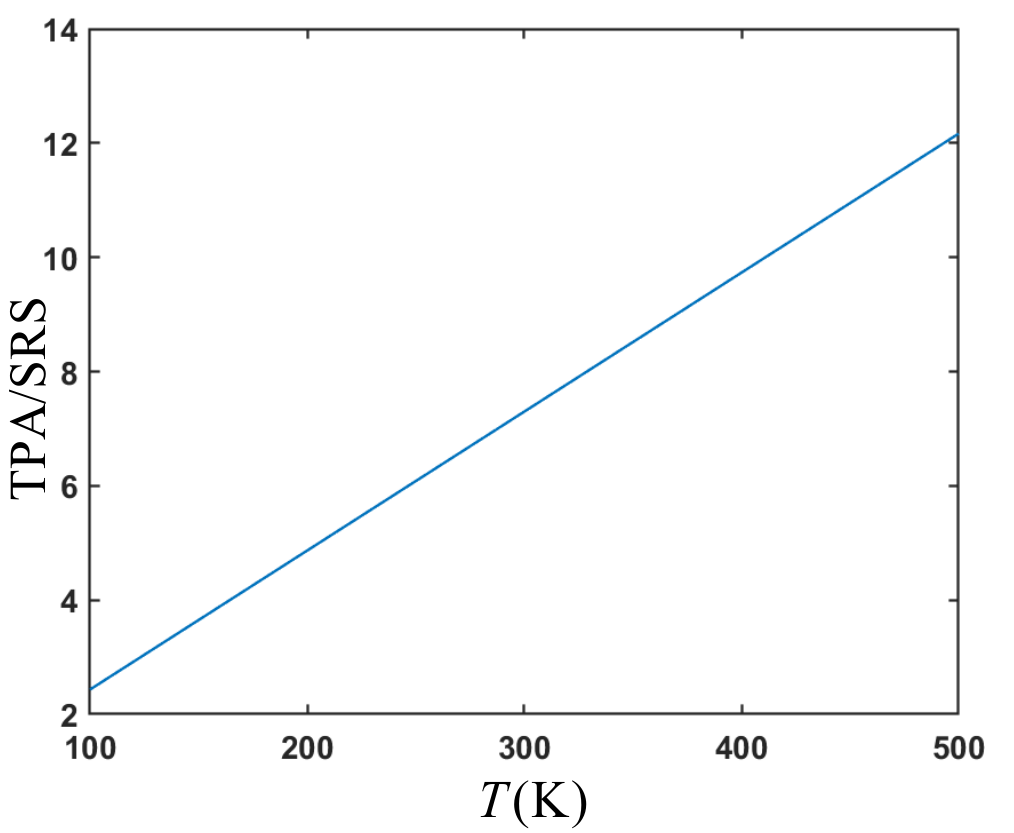}
\caption{The ratio of ETPA and ESRS under their optimal conditions via different temperature.}
\label{fig:re24:temperature}
\end{figure}\par

Furthermore, we examine the influence of temperature, which is intrinsic in Equation~\ref{equ-molecular-SRS-final}. Since the ETPA process occurs over an extremely short timescale, it remains largely unaffected by low-frequency vibrational modes. In contrast, ESRS is sensitive to such modes, and the temperature-dependent parameter $\tilde{D} \propto T$ consequently modulates the signal strength of ESRS. As indicated by Equation~\ref{equ-molecular-SRS-final}, the ESRS signal intensity is approximately inversely proportional to temperature. Therefore, the ratio between the ETPA and ESRS signals varies linearly with temperature, implying that enhanced ESRS performance can be achieved at lower temperatures.

In summary, we have conducted a comparative analysis of ETPA and ESRS using both a three-level atomic model and a more realistic molecular model. These processes represent essential nonlinear optical phenomena with significant implications for spectroscopy, ultrafast dynamics, and biological imaging. For the three-level system, we derived an explicit analytical form to characterize and compare the signal strengths of both processes. Our results indicate that the ratio between ETPA and ESRS signals is primarily determined by the transition dipole moments, which is an intrinsic material property, as well as the total detuning, which depends on both the energy level structure of the sample and the frequency of the input photon pairs. 

To accurately analyze realistic conditions in molecular systems, we introduce a Brownian oscillator model to account for nuclear vibrational dynamics. Numerical simulations demonstrate that ETPA and ESRS can achieve comparable signal intensities under appropriate conditions, such as low vibrational decay rates, frequency-correlated photon pairs, and matched detuning parameters. Moreover, the signal magnitudes of both processes remain within the same order of magnitude across both resonant and off-resonant regimes. Yet ETPA signal is facing some debates caused by low signal-to-noise ratio thus difficult to be detected at low photon flux regime where entanglement is maximized, this may be overcame by using squeezed light with higher flux, improving the detection performance and decrease transmission loss. Above improvements would both affect ETPA and ESRS signal, thus the comparable ratio still valid.

These findings provide a theoretical foundation indicating that both third-order nonlinear processes are experimentally accessible under similar conditions, and that they can be used complementarily to probe material properties. We have also identified key material and experimental parameters that most significantly influence the outcomes, offering practical guidance for future studies. This work suggests that achieving strong ESRS signals are feasible under conditions already demonstrated in ETPA experiments, thereby opening new opportunities for research in ultrafast science and molecular dynamics.

\begin{acknowledgement}

Z. D. Z. and M. Z. gratefully thank the support of  the Excellent Young Scientists Fund by National Science Foundation of China (No.  9240172),  the General Fund by
National Science Foundation of China (No. 12474364), and the National Science Foundation of China/RGC
Collaborative Research Scheme (No. 9054901). F. S. acknowledges support from the Cluster of Excellence "Advanced Imaging of Matter" of the Deutsche Forschungsgemeinschaft (DFG) - EXC 2056 - project ID 390715994, and the research unit "FOR5750: OPTIMAL" - project ID 531215165.

\end{acknowledgement}

\begin{suppinfo}

The following files are available free of charge.
\begin{itemize}
  \item jpc-lett-sup: Derivation of the Cauchy principle value in Eq. 10
\end{itemize}

\end{suppinfo}

\bibliography{ref}

\end{document}